\documentclass[superscriptaddress,twocolumn,prb]{revtex4-1}

\usepackage{graphicx}			% Include figure files
\usepackage{dcolumn}			% Align table columns on decimal point

\usepackage{verbatim}
\usepackage{xcolor}
\usepackage{lipsum}

\usepackage{bm}					% bold math
\usepackage[mathlines]{lineno}	% Enable numbering of text and display math
\usepackage{amsmath, amsthm, amscd, amssymb} % the AMS packages:
\usepackage{enumerate}
\usepackage{romannum}

\usepackage{natbib}
\newcommand{\e}[1]{\mathrm{e}^{#1}}

\newcommand{\unit}[1]{~\mathrm{#1}}
\newcommand{\gray}[1]{\textcolor{gray}{#1}}
\newcommand{\ccal}{c_{\rm cal}}
\newcommand{\odr}{\omega_{\rm dr}}
\newcommand{\Oo}{\Omega_{\rm 0}}
\def \dif {\mathrm{d}}

\begin{document}

\title{{\bf Accurate mass measurement of a levitated nanomechanical resonator for precision force-sensing}}%

\author{F. Ricci}%
\affiliation{ICFO-Institut de Ciencies Fotoniques, The Barcelona Institute of Science and Technology, 08860 Castelldefels (Barcelona), Spain}

\author{M. T. Cuairan}%
\affiliation{ICFO-Institut de Ciencies Fotoniques, The Barcelona Institute of Science and Technology, 08860 Castelldefels (Barcelona), Spain}

\author{G. P. Conangla}%
\affiliation{ICFO-Institut de Ciencies Fotoniques, The Barcelona Institute of Science and Technology, 08860 Castelldefels (Barcelona), Spain}

\author{A. W. Schell}%
\affiliation{ICFO-Institut de Ciencies Fotoniques, The Barcelona Institute of Science and Technology, 08860 Castelldefels (Barcelona), Spain}
\affiliation{Central European Institute of Technology, Brno University of Technology, Purkynova 123, CZ-612 00 Brno, Czech Republic}

\author{R. Quidant}%
\affiliation{ICFO-Institut de Ciencies Fotoniques, The Barcelona Institute of Science and Technology, 08860 Castelldefels (Barcelona), Spain}
\affiliation{ICREA-Instituci\'o Catalana de Recerca i Estudis Avan¸cats, 08010 Barcelona, Spain}

\begin{abstract}
\noindent Nanomechanical resonators are widely operated as force and  mass sensors with sensitivities in the zepto-Newton ($10^{-21}$) and yocto-gram ($10^{-24}$) regime, respectively. Their accuracy, however, is usually undermined by high uncertainties in the effective mass of the system, whose estimation is a non-trivial task. This critical issue can be addressed in levitodynamics, where the nanoresonator typically consists of a single silica nanoparticle of well-defined mass 
Yet, current methods assess the mass of the levitated nanoparticles with uncertainties up to a few tens of percent, 
therefore preventing to achieve unprecedented sensing performances.
Here, we present a novel measurement protocol that uses the electric field from a surrounding plate capacitor to directly drive a charged optically levitated particle in moderate vacuum. The developed technique estimates the mass within a statistical error below $1\%$ and a systematic error of $\sim 2\%$, and paves the way toward more reliable sensing and metrology applications of levitodynamics systems.\\\\
%\noindent {\bf Keywords:} mechanical resonators, optical levitodynamics, levitation optomechanics, \mbox{force~sensing},vacuum
\end{abstract}

\maketitle

%\noindent{\bf Mass measurement of a levitated nano-sensor within 3\% accuracy}\\\\%
%{\bf High-accuracy mass measurement of a levitated nano-sensor}\\\\%

%\noindent{\bf Keywords}\mbox{}\\
%mechanical resonators, optical levitodynamics, levitation optomechanics, force sensing,vacuum

\noindent{\bf Introduction}\mbox{}\\
Nanomechanical resonators play a leading role in the field of force \citep{Moser2013Ultrasensitive}, mass \citep{Chaste2012ANanomechanical}, and charge \citep{Cleland1998ANanometer} sensing. Thermal noise represents the ultimate limitation in the their sensitivity \citep{Yin2013Optomechanics,Norte2016Mechanical}, and hence clamped resonators are usually operated in cryogenic environments \citep{Purdy2012Cavity}.

Owing to their unprecedented decoupling from the environment, levitated nanomechanical systems have recently been able to reach room temperature performances comparable to such clamped cryogenic nanoresonators\citep{Gieseler2012Subkelvin, Jain2016Direct,Setter2018Real-Time}, yet with a sensible reduction of the complexity of the apparatus. Moreover, the negligible mechanical stresses introduced by levitation allow to fulfill the rigid body approximation. As a result, the mass of the resonator is uniquely defined by the inertial mass of the levitated nanoparticle and does not require precise assessment of the system's geometry, knowledge on material properties and complex flexural models for the shape of the oscillation modes, as it is the case for clamped systems.

Despite zepto-Newton resonant force sensitivities with levitated nanoparticles in vacuum have been predicted\citep{Gieseler2013Thermal} and demonstrated~\citep{Ranjit2016Zeptonewton}, and recent experiments with free falling nanoparticles enable for the detection of static forces~\citep{Hebestreit2018Sensing}, the accuracy of these results does not outperform that of existing systems.
%no significative 
% efforts have been done in boosting the accuracy of such measurements.
In most of the levitation experiments, in fact, the uncertainties on the detected forces are of the order of few tens of percent~\citep{Hebestreit2018Sensing}, sometimes even as high as $50\%$~\citep{Hempston2017Force}. Such large errors arise from uncertainties in the particle displacement calibration~\citep{Hebestreit2017Calibration}, whose accuracy is critically affected by the poor knowledge on the particle's mass. This results in severe limitations on their sensing and metrology applications, where the accuracy of a measurement is just as important as its precision.

Silica micro and nano-spheres are the most commonly used type of particle in levitated sensing experiments. Due to their fabrication process~\citep{Stober1968Controlled}, these particles feature a finite size distribution with a $2$--$5\%$ coefficient of variation \citep{microparticles2018Data}. This, together with even higher uncertainties on the density of the amorphous silica used (up to $20\%$~\citep{Parnell2016Porosity}), leads to inaccurate values of the particle's mass. One could avoid assumptions of the manufacturer specifications by relying on the kinetic theory of gasses to calculate the radius of the particle~\citep{Beresnev1990Motion}. Also in this case, however,  the final measurement of the mass is affected by uncertainties on the material density and on other quantities, such as pressure and molar mass~\citep{Hebestreit2018Measuring} of the surrounding media. A more accurate estimation of the particle's mass is therefore highly desirable, as it would boost the accuracy of sensors based on levitated particles.

Here, we propose and experimentally demonstrate a measurement protocol that is unaffected by the above-mentioned uncertainties (density, pressure, size, etc.), and leads to an assessment of the particle's mass within $2.2\%$ systematic error and $0.9\%$ statistical error. Our method exploits a new design of an optical trap in which a pair of electrodes is placed around the focus and is based on the analysis of the response of a trapped charged particle to an external electric field. Careful error estimation has been carried out in order to assess the final mass uncertainty, including the treatment of possible anharmonicities in the trapping potential. The technique we propose is easy to implement in any vacuum trapping set-up and improves by more than an order of magnitude the accuracy of most precision measurements.

\begin{figure}
\includegraphics[width=\linewidth]{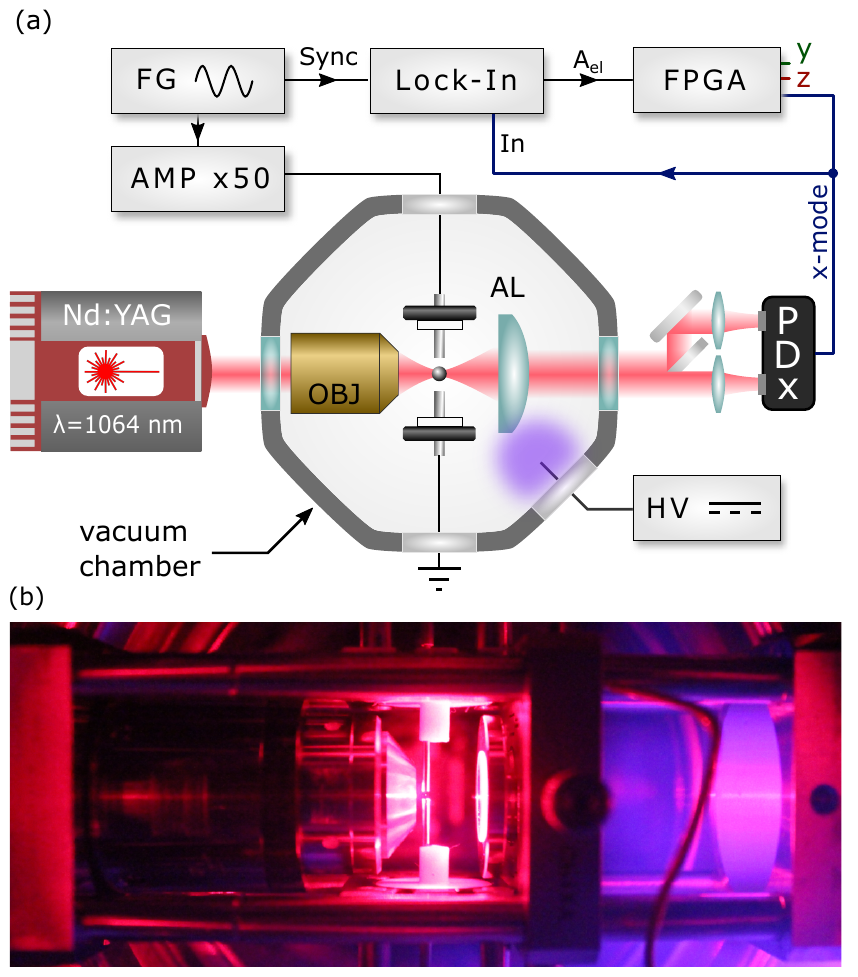}
\caption{\footnotesize \textbf{Experimental set-up} (\textbf{a}) A microscope objective (OBJ) focuses a laser beam inside a vacuum chamber, where a single
silica nanoparticle is trapped in the focus. The light scattered by the particle is collected with an aspheric lens (AL) and the motion of the particle is detected in a split detection scheme. A pair of electrodes is connected to the amplified signal from a function generator (FG), creating an electric field that drives the charged particle. An FPGA and a lock-in amplifier are used to bandpass and record the signal from the detector. (\textbf{b}). A camera image of the set-up inside of the vacuum chamber. The purple glow on the side of the chamber is emitted by a  generated by a bare electrode connected to a high voltage (HV) DC source and is used to control the net charge of the particle.}
\label{fig:01_Exp_SetUp}
\end{figure}

\vspace{0.3cm}
%\clearpage
\noindent{\bf Experimental Set-Up}\mbox{}\\
The experimental set-up is depicted in Fig.~\ref{fig:01_Exp_SetUp}a. A single silica nanoparticle ($d=143 \pm 4 \unit{nm}$ in diameter; nominal value of the manufacturer) is optically trapped in vacuum with a tightly focused laser beam (wavelength $\lambda = 1064\unit{nm}$, power $P\simeq 75\unit{mW}$, $N\!A=0.8$). The oscillation of the particle along the $x$--mode is monitored with a balanced split detection scheme that provides a signal $v(t)$ proportional to the particle displacement $x(t)=v(t) / \ccal $, with $\ccal$ being the linear calibration factor of the detection system\citep{Hebestreit2017Calibration}. Along the same axis, a pair of electrodes (see Fig.~\ref{fig:01_Exp_SetUp}b and inset) form a parallel plate capacitor that we use to generate an oscillating electric field $E(t)=E_0\cos( \odr t )$ at the particle position, which in turn induces a harmonic force $F_{\rm el}(t)$ on the charged particle.

\noindent The equation of motion of the particle can be described  by a thermally and harmonically driven damped resonator:
\begin{equation} \label{eq:motion_02}
m \ddot x  +  m \Gamma \dot{x} +  kx = F_{\rm th}(t) + F_{\rm el}(t)~.
\end{equation} 
Here, $m$ is the mass of the particle, $\Gamma$ is the damping rate and $k=m\Omega_0^2$ is the stiffness of the optical trap, with $\Omega_0$ being the mechanical eigenfrequency of the oscillator. The first forcing term $F_{\rm th}$ models the random collisions with residual air molecules in the chamber. It can be expressed as $F_{\rm th} = \sigma \eta(t)$, where $\eta(t)$ has a Gaussian probability distribution that satisfies $\langle \eta(t) \eta(t+t') \rangle = \delta(t') $, and $\sigma$  relates to the damping via the fluctuation-dissipation theorem: $\sigma  = \sqrt{2 k_B T m \Gamma}$, with $k_{\rm B}$ being the Boltzmann constant and $T$ the bath temperature. The second forcing term $F_{\rm el}$ arises from the Coulomb interaction of the charged particle with the external electric field $E(t)$, and can be expressed as $F_{\rm el}(t) = F_0 \cos(\odr t)$, where $F_0 = q \cdot E_0$. The net charge $q=n_q \cdot q_e$, with $q_e$ being the elementary charge and $n_q$ the number of charges on the particle, can be  controlled~\citep{Frimmer2017Controlling} by applying a high DC voltage $V_{\rm HV} \sim \pm 1\unit{kV}$ on a bare electrode placed on a side of the vacuum chamber. Via the process of corona discharge~\citep{Fridman2004Plasma}, this creates a plasma consisting of a mixture of positive or negative ions (depending on the $V_{\rm HV}$ polarity) %$(mainly $N_2^{+}$ and $O_2^{+}$)
 and electrons %$e^{-}$
  that can ultimately add to, or remove from, the levitated particle one single elementary charge at a time. Positive and negative ions are accelerated towards opposite directions due to the presence of the electric field from the electrode. As a result, the ratio of positive to negative charges reaching the particle is biased by the electrode polarity, thus allowing us to fully control the final charge of the particle within positive or negative values (see Supplementary Section~S3).
This is a significant advantage compared to other discharging techniques that rely on shining UV light on the particle~\citep{Moore2014Search}, where the net charge can only be diminished until reaching neutrality. \\\\
%
%\vspace{0.3cm}
\noindent{\bf Measurement}\mbox{}\\
A single nanoparticle is loaded in the trap at ambient pressure by nebulizing a solution of ethanol and monodispersed silica particles into the chamber. The pressure is then decreased down to $P \lesssim 1\unit{mBar}$ where the net number of charges $n_q$ can be  set with zero uncertainty. Finally, the system is brought back up to an operating pressure of $P\simeq 50\unit{mBar}$. At this pressure the particle is in the ballistic regime, but its dynamics is still highly damped. This condition is favorable for our experiments, as the high damping reduces the contribution of anharmonicities to the dynamics of the particle~\citep{Gieseler2013Thermal}, allowing  us to apply the fully linear harmonic oscillator model which predicts:
\begin{align}
& S_x(\omega) = S_{x}^{\rm th}(\omega) + S_x^{\rm el}(\omega) \nonumber \\
& = \frac{4 ~ k_{\rm B} T ~\Gamma}{m\left[ \left( \omega^2 - \Omega_0^2 \right)^2 + \Gamma^2 \omega^2 \right]} + \frac{F_0^2 ~ \tau ~ {\rm sinc}^2 [(\omega-\odr)\tau]}{m^2 \left[ \left( \omega^2 - \Omega_0^2 \right)^2 + \Gamma^2 \omega^2 \right]}. \label{eq:PSD01}
\end{align}
Here, $S_x(\omega)$ is the single-sided Power Spectral Density (PSD) of the thermally and harmonically driven resonator, whose dynamics $x(t)$ is being observed for a time $\mathcal{T}=2\tau$.
%while $\chi^2(\omega) = m^{-2} \left[ \left( \omega^2 - \Omega_0^2 \right)^2 + \Gamma^2\omega^2 \right]^{-1}$ is the squared susceptibility of the harmonic oscillator.
Note that $S_x(\omega)$ relates to the experimentally measured PSD $S_v(\omega)$ via the calibration factor $\ccal$, such that $S_v(\omega) = \ccal^2 \cdot S_x({\omega})$~\citep{Hebestreit2017Calibration}.

In the absence of the electric driving, the motion of the particle in the optical trap is purely thermal and its PSD is well approximated by a typical Lorentzian function.
From an experimental measurement of $S_v^{\rm th}(\omega)$ we can extract the value of $S_v^{\rm th}(\odr)$ and perform maximum likelihood estimation (MLE) to obtain the values of $\Omega_0$ and $\Gamma$ as fitting parameters. Likewise, when the coherent driving is applied to the system, we are able to determine the magnitude of the driven resonance $S_v(\odr)$ and to calculate from this measurement the solely electric contribution  $S_v^{\rm el}(\odr) = S_v(\odr) - S_v^{\rm th}(\odr)$. 
\begin{figure}
\includegraphics[width=\linewidth]{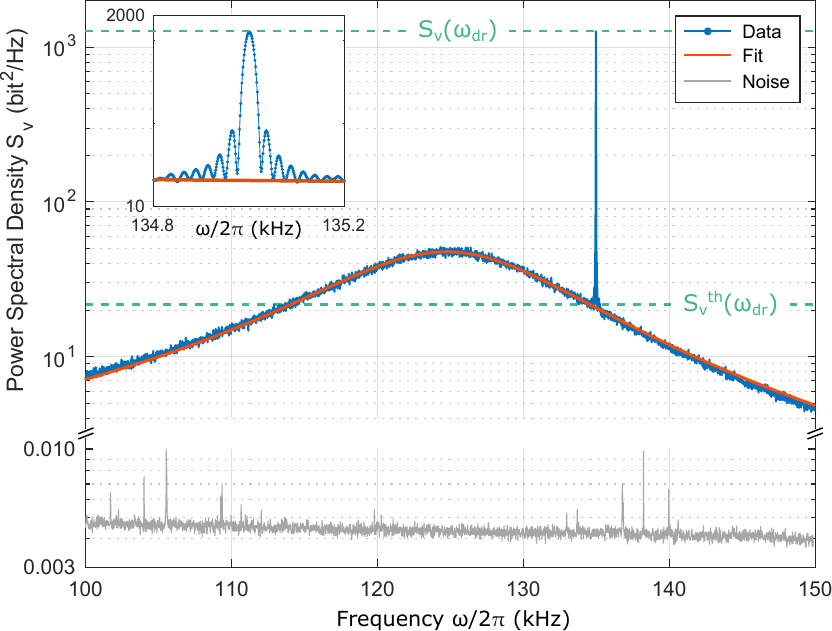}
\caption{\footnotesize \textbf{Measurement.} Power Spectral Density $S_v(\omega)$ of a thermally and harmonically driven resonator at $P=50\unit{mBar}$. The broad peak centered at $\Oo \simeq 125\unit{kHz}$ is the oscillator response to the thermal driving, that we fit with a Lorentzian function (orange) to extract $S_v^{\rm th }(\odr)$, together with $\Oo$, $\Gamma$ and the corresponding uncertainties. The narrowband peak at $\omega=135\unit{kHz}$, also shown in detail in the inset, depicts to the electric excitation, from which we retrieve $S_v^{\rm el}(\odr) = S_v(\odr)-S_v^{\rm th}(\odr) $. The gray spectrum at the bottom of the plot is the measurement noise, which is $\gtrsim 40\unit{dB}$ below the particle's signal.}
\label{fig:02_Measurement}
\end{figure}
Figure~\ref{fig:02_Measurement} exemplifies this process for  $\odr/(2\pi) = 135\unit{kHz}$ and for a signal-to-noise ${\rm SNR} = S_v(\odr) / S_v^{\rm th}(\odr) \simeq 60$. The curve shown is computed with Bartlett's method from an ensemble of $N_{\rm psd} = 1000$ averages of individual PSDs, calculated from $\mathcal{T}=40\unit{ms}$ position time traces. In Supplementary Section~S2 we verify that over the whole measurement time $t=N_{\rm psd} \times \mathcal{T} = 40\unit{s}$ the system does not suffer from low frequency drifts. The electrically driven peak can be fully resolved (see inset in Fig.~\ref{fig:02_Measurement}), and its shape agrees with the Fourier transform of the rectangular window function used for PSD estimation. The gray trace at the bottom of the plot represents the measurement noise, which is $\sim 40\unit{db}$ below the thermal signal and more than $55\unit{db}$ below the driven peak. Finally, the solid line is a MLE fit of a thermally driven Lorentzian to the experimental data. 
Note that, to perform the fit and to retrieve the value of $S_v^{\rm th}(\odr)$, the electrically driven peak is numerically filtered out by applying to the time series data-set a notch filter of variable bandwidth $b$ around $\odr$. The value of $b$ depends on the driving amplitude, with typical values of the order of tens of Hertz. In Supplementary Section~S7 we show how this method introduces negligible errors that remain always below $\sim 0.01~\%$. 

The mass of the particle can ultimately be calculated considering the ratio $R_S=\frac{S_v^{\rm el} (\odr)}{S_v^{\rm th} (\odr)} =  \left. \frac{S_v-S_v^{\rm th}}{S_v^{\rm th}}\right|_{\omega=\odr}$. In fact, note that while both $S_v^{\rm el} $  and $S_v^{\rm th}$ depend quadratically on $\ccal$, the latter scales as $m^{-1}$ while the former scales as $m^{-2}$. Thus, from their ratio we obtain:
\begin{equation} \label{eq:mass}
m = \frac{n_q^2 ~q_e^2 ~ E_0^2 ~ \mathcal{T} ~ }{8~ k_{\rm B} T ~ \Gamma ~ R_S}~.
\end{equation}

To ensure the validity of the linear resonator model, we also considered a cubic term in the restoring force and performed Montecarlo Simulations of the resulting Duffing resonator with parameters compatible with our experimental settings and an overestimated value of the Duffing coefficient~\citep{Ricci2017Optically,Gieseler2014Nonlinear} $\xi = 12\unit{\mu m}^{-2}$. The outcome of the simulations is detailed in Supplementary Section~S5, and confirms the negligibility of the nonlinear terms for pressures of $P\simeq 50\unit{mBar}$. We stress that this assumption fails already at slightly lower pressures of $\sim 10\unit{mBar}$ where a more complicated non-linear response model would be needed. 

\begin{figure}[h]
\vspace{15pt}
\includegraphics[width=\linewidth]{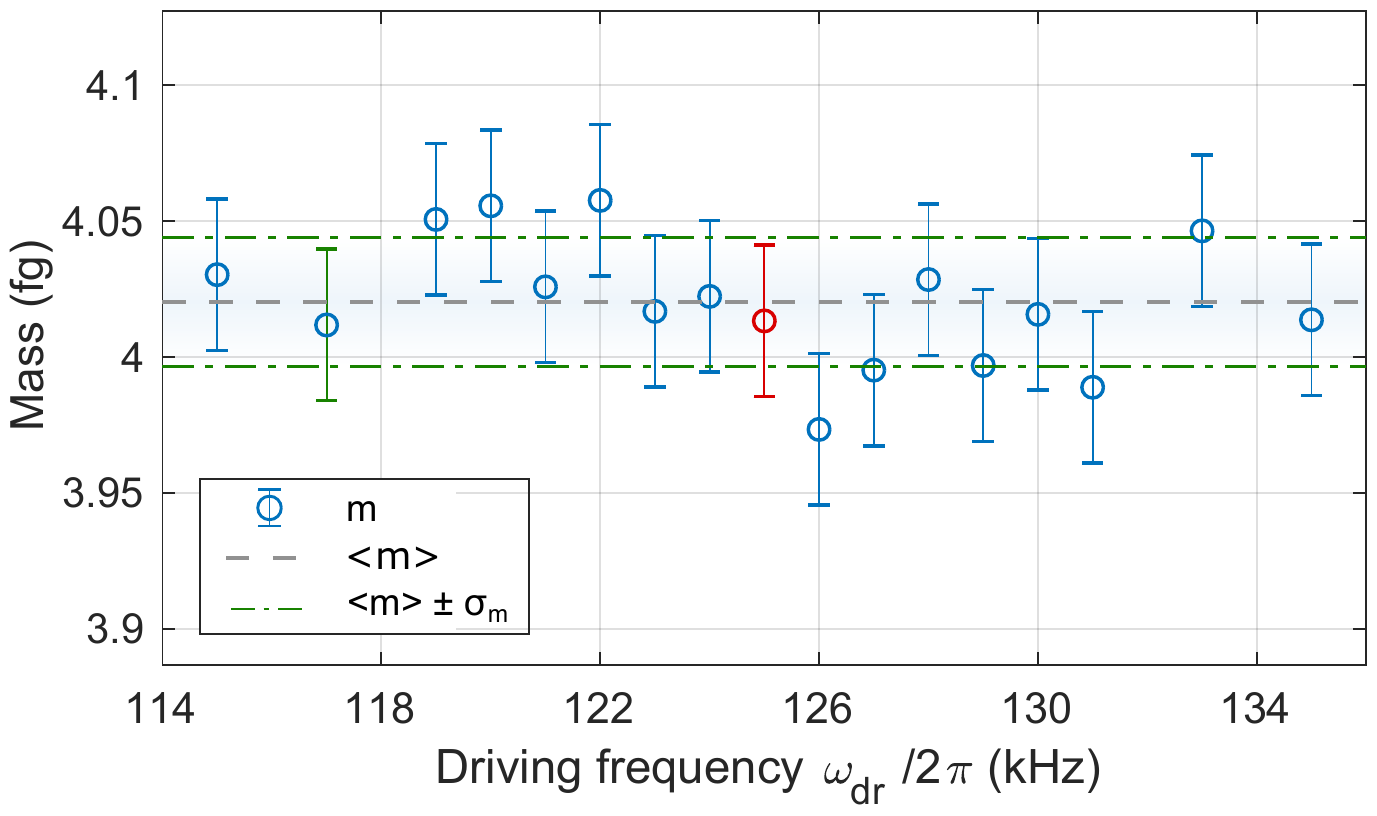}
\caption{\footnotesize \textbf{Results of mass calculation.} (\textbf{a}), The mass of the levitated nanoparticle $m$ is calculated for different driving frequencies $\odr$. Error bars correspond to the statistical error $\sigma_{m}^{\rm stat}$, calculated from a reproducibility measurement of $20$ independent datasets of same experimental conditions: $\odr = 125\unit{kHz}$ and ${\rm SNR} = 60$. The measurement is shown to be independent from the chosen driving frequency $ \odr$, and the standard deviation (green horizontal lines) displays compatibility with the statistical error. Values reported in Table~\ref{tab:Errors} correspond the the red-highlighted point at $\odr=125kHz$.} 
\label{fig:03_Results}
\end{figure}

\vspace{0.3cm}
\noindent{\bf Error Estimation}\mbox{}\\
In order to estimate the systematic error committed in calculating the mass, a careful study of all the sources of error has to be carried out. Table~\ref{tab:Errors} summarizes the absolute values and the relative uncertainties of the quantities entering in Eq.~\eqref{eq:mass}. The specific case reported corresponds to point at $\odr = 125\unit{kHz}$ of the data shown in Fig.~\ref{fig:03_Results}.
%As shown Supplementary Section \red{???}, the number of elementary charges can be exactly measured, such that $\sigma_{n_q} = 0$.
For several variables and constants, we can neglect the corresponding uncertainty. Accordingly, for the error propagation we set: $\sigma_{q_{e}} = \sigma_{\mathcal{T}} = \sigma_{k_{\rm B}} =0 $. Note that the specific number of charges $n_q = 8$ chosen in our measurements is arbitrary. Other measurements have been previously carried out with different values of $n_q$ and have confirmed the independency of the method from $n_q$, provided the dynamics is maintained in the linear regime of oscillation. Concerning the other quantities, instead, we follow the arguments stated below: 
\begin{enumerate}[(i)]
\item The electric field was simulated with the \emph{finite elements method}, and was mainly affected by uncertainties in the geometry of the electrodes
%which was measured with a portable microscope up to a $10\unit{\mu m }$ precision.
(see Supplementary section~S4 for further details). We measure a distance between electrodes of $d_{\rm el} = 1410\unit{\mu m} \pm 13\unit{\mu m}$, and a corresponding electric field (for an applied dc potential of $1V$) $E_0= 577 \pm 6\unit{V/m}$.
\item The two heights of the power spectral densities $S_v(\odr)$ and $S_v^{\rm th}(\odr)$ from which the ratio $R_S$ is calculated are only affected by statistical errors since simulations confirm the validity of the linear model. $\sigma_{S_v}$ is thus calculated from an ensemble of $N_{\rm psd}$ measurements as the standard error of the mean, with the $1/\sqrt{N_{\rm psd} }$ trend being verified. The same applies for $S_v^{\rm th}$, where in this case $\sigma_{S_v^{\rm th}}$ is calculated in the absence of external electric driving. 
\item The thermal bath surrounding the partcle is assumed to be constantly thermalized with the set-up, and more precisely with the vacuum chamber walls. Again, the moderate-high pressure $P= 50\unit{mBar}$ ensures this assumption. Multiple temperature measurements on the surface of the vacuum chamber are carried out with a precision thermistor ($0.5^{\circ}{\rm C}$ accuracy) in order to exclude the presence of temperature gradients  and significant variations during the experimental times (see Supplementary Section~S8 for data and further discussion).
\item The uncertainty of fitting parameters such as $\Omega_0$ and $\Gamma$ can be extracted directly from the lorentzian fits.
\end{enumerate}

The variables involved in Eq.~\ref{eq:mass} can be considered uncorrelated and the standard uncertainty propagation\citep{Ku1966Notes}  can be performed. A detailed derivation is provided in Supplementary Section~S9.
\mbox{}\\
\begin{table}
\begin{center}
\footnotesize{
  \begin{tabular}{ c | c | c } 
    \hline
    Quantity & Value $ z_i $ & Error $\sigma_{z_i}/z_i$ \\
    \hline
    \hline
    \gray{$n_q$}~ 							&~ \gray{$8$}	~ 										&~ \gray{$0$}~ 					\\ 
    \gray{$q_e$}~ 							&~ \gray{$1.602\times 10^{-19}\unit{C}$}~ 				&~ \gray{$6.1\times 10^{-9}$} 	\\ 
	$E_0$~ 									&~ $5.305 \unit{kV m^{-1}}$~ 							&~ $0.011$ 						\\ 
	\gray{$\mathcal{T}$}~ 							&~ \gray{$40\unit{ms}$}~ 								&~ \gray{$4 \times 10^{-5}$} 	\\ 
    $S_v(\odr)$ 							&~ $1057.8\unit{bit^2 Hz^{-1}} $~ 						&~ $ 0.005$ 					\\ 
    $S_v^{\rm th}(\odr)$ 					&~ $14.7 $~ 												&~ $0.007$ 						\\
    \gray{$k_{\rm B} $} 					&~ \gray{$1.380 \times 10^{-23}\unit{J K^-1} $}~		&~ \gray{$5.72 \times 10^{-7}$} \\ 
    $T$ 									&~ $295.8\unit{K}$~										&~ $ 0.002$ 					\\ 
   	$\Gamma$ 								&~ $1.998 \times 10^5\unit{rad s^{-1}}$~				&~ $ 0.003$ 					\\ 
   	\hline
   $\mathbf{m}$										&~ $\mathbf{4.01}$~\bf{fg}~									&~ $\mathbf{0.009}~(\rm stat.) \pm \mathbf{0.024}~(\rm syst.)$ \\
   	\hline
    \end{tabular}
  }
\end{center}
\caption{\footnotesize \textbf{Uncertainties table.} The different quantities $z_{i}$ involved in the calculation of the mass are here reported together with the corresponding error $\sigma_{z_i}$. Color coding indicates negligibility of the uncertainty, with gray rows  implying $\sigma_{z_i} \simeq 0$.  }
\label{tab:Errors}
\end{table}

\vspace{0.2cm}
\noindent{\bf Results}\mbox{}\\
The statistical error $\sigma_{m}^{\rm stat}$ of our measurement is calculated from the standard deviation of a set of $20$ independent measurements performed at $\odr/2\pi = 130\unit{kHz}$ and for ${\rm SNR} \simeq 60$. We find $\sigma_{m}^{\rm stat}/m = 0.7\%$. This dispersion is displayed as error bars in Fig.~\ref{fig:03_Results}, where we plot the calculated mass as a function of $\odr$, again for a ${\rm SNR} \simeq 60$. The region within green dot-dashed lines corresponds to the standard deviation $\pm \sigma_{m}^{\rm sweep}$ of the presented data. 

The compatibility $\sigma_{m}^{\rm stat} \simeq \sigma_{m}^{\rm sweep}$, and the reproducibility of the mass calculated at different driving frequencies reveal that the measurements are not affected by nonlinearities in the system. In fact, strong driving fields lead to anharmonic particle dynamics which in turn introduce an unphysical mass dependency on $\odr$. Fig~ S6a in Supplementary Section~S2 exemplifies this situation and shows how in the non-linear regime the calculated mass is affected by sever systematic errors. In our method we avoid this situation maintaining the driving field amplitude below $5.5\unit{kV/m}$. In this regime, we have additionally verified the independency of the calculated mass from the electric field $E_0$ and tested the quadratic scaling of $R_S$ as a function of $E_0$. These measurements are described in fig.~S3 of the Supplementary Information. The excellent agreement with the model provides a further validation of eq.~\eqref{eq:mass} and of the harmonic approximation made. As a final remark, we compare the measured mass $m$ of a $d=143 \pm 4 \unit{nm}$ diameter particle with the one calculated from the manufacturer specifications $m_{\rm man}$. Assuming a nominal density for St\"ober silica of $\rho_{\rm p}=2200~{\rm kg}/{\rm m}^3$ and propagating the corresponding uncertainties one finds $m_{\rm man} = (3.37 \pm 0.84)~{\rm fg}$, which shows good agreement with the value measured with our method $m = (4.01 \pm 0.1)~{\rm fg}$.

\mbox{}\\
%\vspace{0.3cm}
\noindent{\bf Conclusions}\mbox{}\\
In conclusion, we presented a novel protocol to calculate the mass of a levitated nano-sensor through its electrically driven dynamics. We stress that this method only assumes a driven damped harmonic oscillator. As such, it is suitable to measure the oscillator's mass in a large variety of optical trapping systems and possibly also in more general mechanical resonators schemes. The level of precision and accuracy obtained establishes an improvement of more than one order of magnitude compared to the state-of-the-art methods, enabling paramount advances in the applications of levitated systems as force sensors and accelerometers. Moreover this technique leads to a much more reliable calibration of the particle's displacement~\citep{Hebestreit2017Calibration}, again providing an important step for the use of levitated systems for metrology and sensing applications, and towards compliance requirements of groundbreaking experiments such as MAQRO~\citep{Kaltenbaek2016Macroscopic}. 

%\mbox{}\\
%\vspace{0.3cm}
%\noindent{\bf AUTHOR INFORMATION}\mbox{}\\
\noindent{\bf Corresponding authors}\\
francesco.ricci@icfo.eu\\
romain.quidant@icfo.eu\\
{\bf ORCID}\\
Francesco Ricci: 0000-0002-5971-3369\\
Romain Quidant: 0000-0001-8995-8976\\
{\bf Aurthor Contributions}\\
F.R. and A.S. conceived the experiment. F.R. designed and
implemented the experimental set-up and wrote all data acquisition software. F.R. and M. T. performed the experiment and analysed the data, with inputs from G.P. Montecarlo and COMSOL simulations were performed by G.P. and M.T. respectively. All authors contributed to manuscript writing. R.Q. and A.S. supervised the work.

\mbox{}\\
\vspace{0.3cm}
\noindent{\bf Acknowledgement}\mbox{}\\
We acknowledge financial support from the ERC- QnanoMECA (Grant No. 64790), the Spanish Ministry of Economy and Competitiveness, under grant FIS2016-80293-R and through the `Severo Ochoa' Programme for Centres of Excellence in R\&D (SEV-2015-0522), Fundaci\'o Privada CELLEX and from the CERCA Programme/Generalitat de Catalunya. We also acknowledge  N. Meyer and the rest of the PNO trapping team. F.R. acknowledges Dr. M. Frimmer and Prof. L. Novotny from ETH (Zurich)for valuable discussions and A. Bachtold (ICFO) for providing a general perspective on the accuracy of mechanically resonating nanosensors.
%
%%%%%%%%%%%%%%%%%%%%%%%%%%%%%%%%%%%%%%%%%%%%%%%%%%%%%%%
%%%%%%%%%%%%%%%%% 	BIBLIOGRAPHY   %%%%%%%%%%%%%%%%%%%%
%%%%%%%%%%%%%%%%%%%%%%%%%%%%%%%%%%%%%%%%%%%%%%%%%%%%%%%
%\clearpage
%\bibliographystyle{ieeetr}
%\bibliography{../../../Thesis/tex/00_ThesisBibliography}

\clearpage
%\newpage
%\mbox{}\\
%\newpage
%\vspace{.5cm}

\noindent{\bf SUPPLEMENTARY INFORMATION} \mbox{}\\

\vspace{0.3cm}
\noindent{\bf S1.\hspace{.2cm}Nonlinear contributions} \mbox{}\\
The nonlinearities in the dynamics of the particle arise from the anharmonicity of the optical potential\citep{Gieseler2013Thermal}. The system can then be modeled as a Duffing resonator for which the linear stiffness becomes a function of position $k(x)$ and the equation of motion therefore reads:
\begin{equation} \tag{S1} \label{eq:motion_NL}
m \ddot x  +  m \Gamma \dot{x} +  m\Oo^2\left(1+\xi x^2 \right)x = \mathcal{F}_{\rm th}(t) + F_{\rm el}(t)~,
\end{equation}
where the nonlinear coefficient $\xi<0$ is the so called \emph{Duffing} coefficient. 

Prompt consequence of the presence of nonlinearities in the dynamics of a resonator is that the eigenfrequency $\Omega_0$ does not correspond to the curvature of the harmonic potential, but gets shifted (down-shifted if $\xi<0$, up-shifted if $\xi>0$) and becomes energy dependent. More precisely we have:
\begin{equation} \tag{S2} \label{eq:S_NLFreqShift}
\Omega_{\rm NL} = \Omega_0 \left( 1+ \frac{3}{4} \xi \langle x^2 \rangle \right), 
\end{equation}
where $\langle x^2 \rangle$ is the variance of the oscillation.

\begin{figure}
\includegraphics[width=\linewidth]{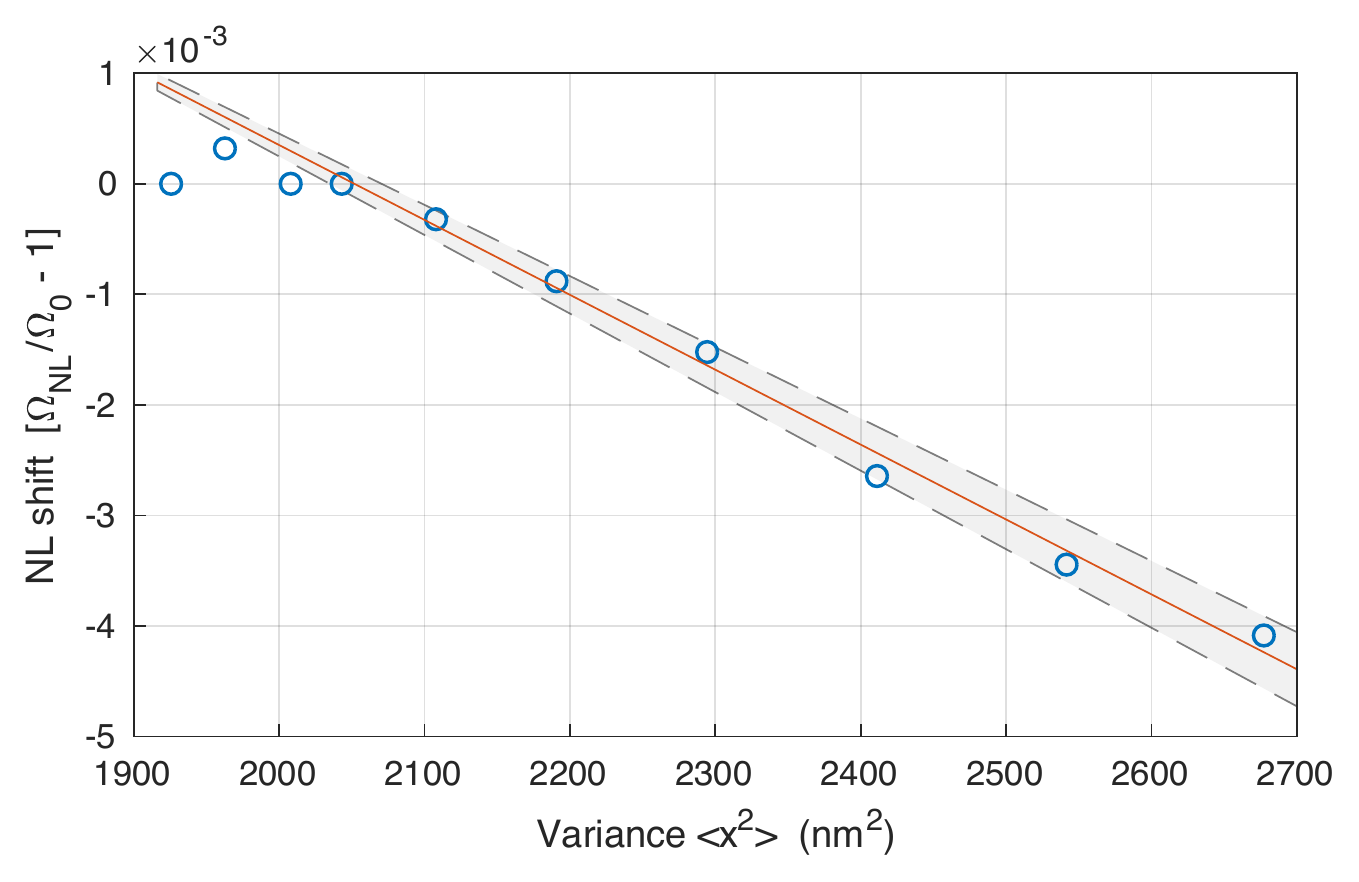}
\caption{\footnotesize \textbf{Nonlinear frequency shift.} Due to the anharmonicity of the optical potential, a nonlinear frequency shift is observed when the particle is excited by the electric actuation. For variances $\langle x^2 \rangle \gtrsim 2000\unit{nm^2}$ we observe a linear shift in agreement with Eq.~\eqref{eq:S_NLFreqShift}. A linear regression (red solid line) and related standard deviation (gray area) allow to extract the value of the nonlinear coefficient $\xi = (-9.03 \pm 0.44)\unit{\mu m ^{-2}}$.}
\label{fig:S03_PlotNLFreqShift}
\end{figure}

We can exploit equation~\eqref{eq:motion_NL} to retrieve the value of $\xi$ by driving the particle with increasingly stronger electric field $E_0$ and monitoring how the frequency  gets shifted by the non-linearities. We use the calibration factor $c_{\rm cal}$ computed with the smallest driving (for which no shift is observed, i.e. $\Omega_{\rm NL} = \Oo$) to convert the experimental variance $\langle v^2 \rangle$ into a calibrated $\langle x^2 \rangle$ with physical units of ${\rm nm^2}$. Figure~\ref{fig:S03_PlotNLFreqShift} shows the nonlinear frequency shift $\frac{\Omega_{\rm NL}}{\Oo}-1$ for increasing variance $\langle x^2 \rangle$. Low driving does not affect the energy of the particle, and no shift is indeed observed. For $\langle x^2 \rangle \gtrsim 2100\unit{nm^2}$ a shift in frequency is observed and we perform a linear fit to the experimental data data according to Eq.~\eqref{eq:S_NLFreqShift}. From the fit we retrieve: $\xi = (-9.03 \pm 0.44)\unit{\mu m ^{-2}}$. The obtained value is in agreement with previous measurement of $\xi$ obtained from completely different methods~\citep{Ricci2017Optically,Gieseler2014Nonlinear}

\begin{figure}
\includegraphics[width=\linewidth]{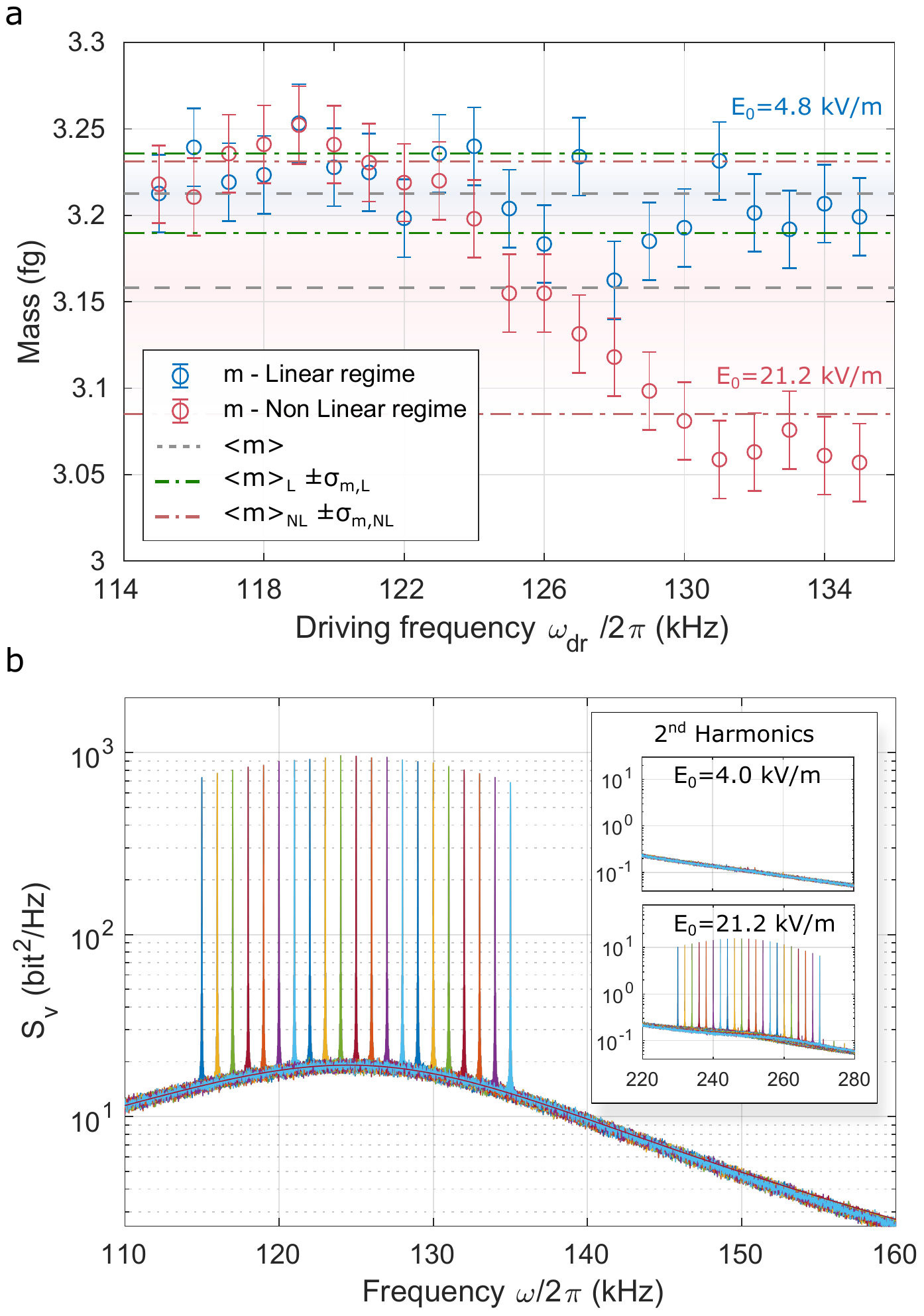}
\caption{\footnotesize \textbf{Nonlinearity based artefacts on mass measurement.} (a) Mass measurement datasets for different driving amplitudes. Blue data  correspond to a harmonic oscillator, where no dependency on the driving frequency is observed. Red data, instead, are measured with the resonator driven into the non-linear regime, and exhibit systematic artefacts depending on the driving frequency. (b) Nonlinear features such as second harmonic oscillations are detected when the resonator is driven in the non-linear regime.}
\label{fig:S03_MassMeas}
\end{figure}

The effect of non-linearities is also tested and characterized on the mass measurement protocol. Figure~3 in the main text presents the mass measured using different driving frequencies $\odr$. Measured data exhibits a standard error $\sigma_{m}^{\rm sweep}$ (shown as green dot-dashed line) that matches the expected statistical error $\sigma_{m}^{\rm stat}$ (given by the errorbar size). This observation is further supported by the linearity of the dynamics and demonstrates the general applicability of the method independently from the chosen driving frequency $\odr$. However, the use of the linear model to fit a Duffing resonator introduces appreciable systematic errors in the analysis: a shift between the mean values is observed and the measured mass now carries an \emph{unphysical} dependency on the driving frequency. This artefact is clearly shown in Fig.~\ref{fig:S03_MassMeas}a, where we compare the results of our protocol for different driving strengths. Blue data corresponds to a resonator in the linear regime, driven by a field amplitude $E_0 = 4.8\unit{kV/m}$, while red data correspond the anharmonic oscillator with driving amplitude $E_0 = 21.2\unit{kV/m}$. As a result, throughout all our measurements we maintain the driving field amplitude below $ 5.5\unit{kV/m}$.
Figure~\ref{fig:S03_MassMeas}b presents the response of the system to the external excitation when the driving frequency is swept in the range $\odr \in [115,135]\unit{kHz}$. For the main figure a driving voltage of $V_{\rm dr}^{\rm pp}=13.7\unit{V}$ was used, resulting in an electric field amplitude $E_0 = 4.0\unit{kV/m}$. For such amplitudes the particle is still driven in the linear regime. The presence of nonlinearities for higher drivings is exemplified in the insets of Fig.~\ref{fig:S03_MassMeas}b, where second harmonic oscillatory components are detected only for $E_0 \gtrapprox 6\unit{kV/m}$.
\begin{figure}
\includegraphics[width=\linewidth]{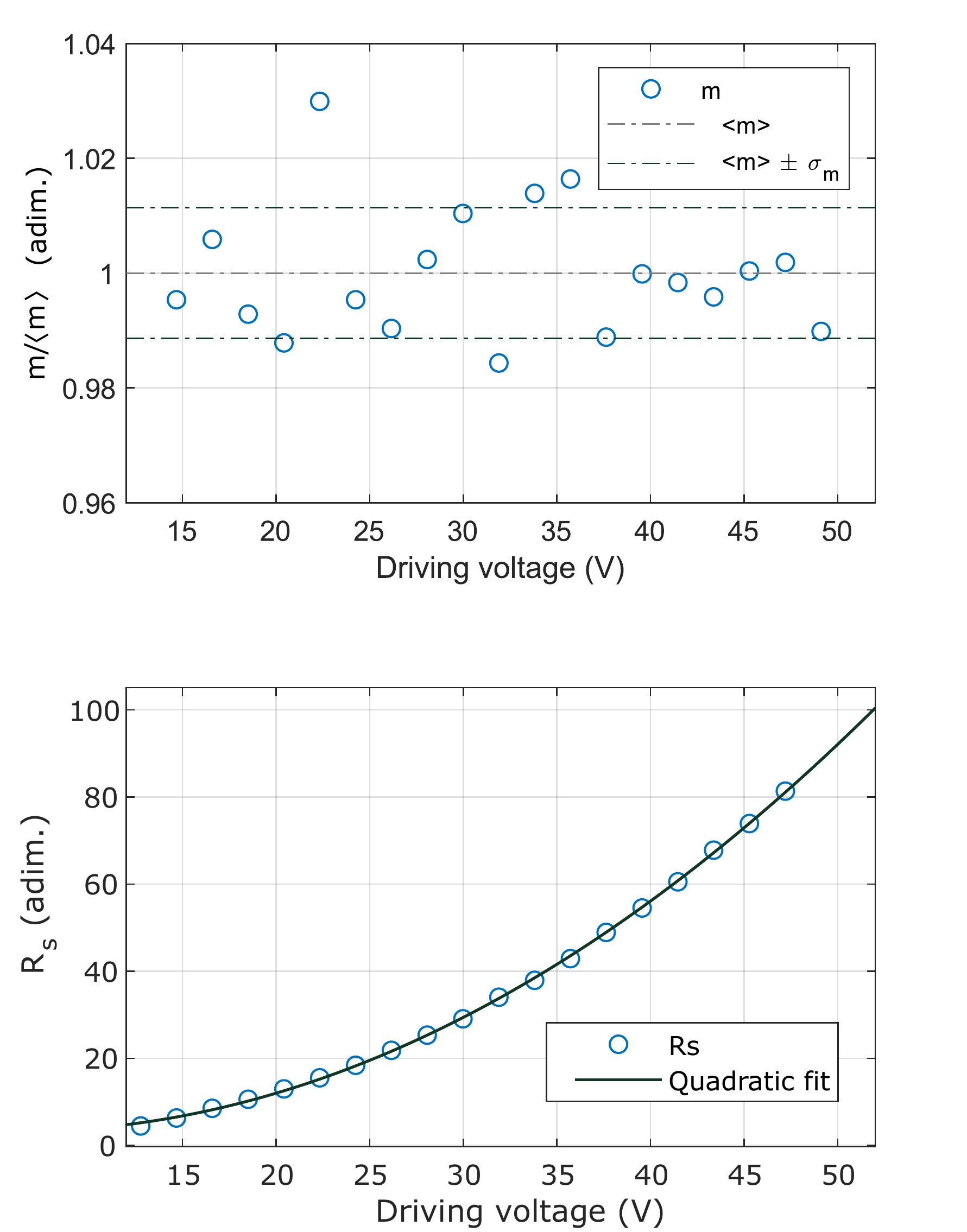}
\caption{\footnotesize \textbf{Independency of measured mass from driving field.} (a) To verify the model and the harmonic approximation made in our study, the mass is measured at varied values of driving voltage $V_{\rm dr} \propto E_0$ below the onset of nonlinearities. Here we measure a bigger particle compared to main text (nominal diameter $d_{\rm p}=235~{\rm nm}$) and the results are shown in relative units normalized to the mean value $\langle m \rangle$ in order to show that the statistical error is maintained below $1\%$ also in this case. (b) The scaling of $R_S$ as a function of $V_{\rm dr}$. The black solid line is a quadratic fit, in agreement with eq.\eqref{eq:PSD01}.}
\label{fig:S03_MassVsDriving}
\end{figure}

\begin{figure}[h]
\includegraphics[width=0.9\linewidth]{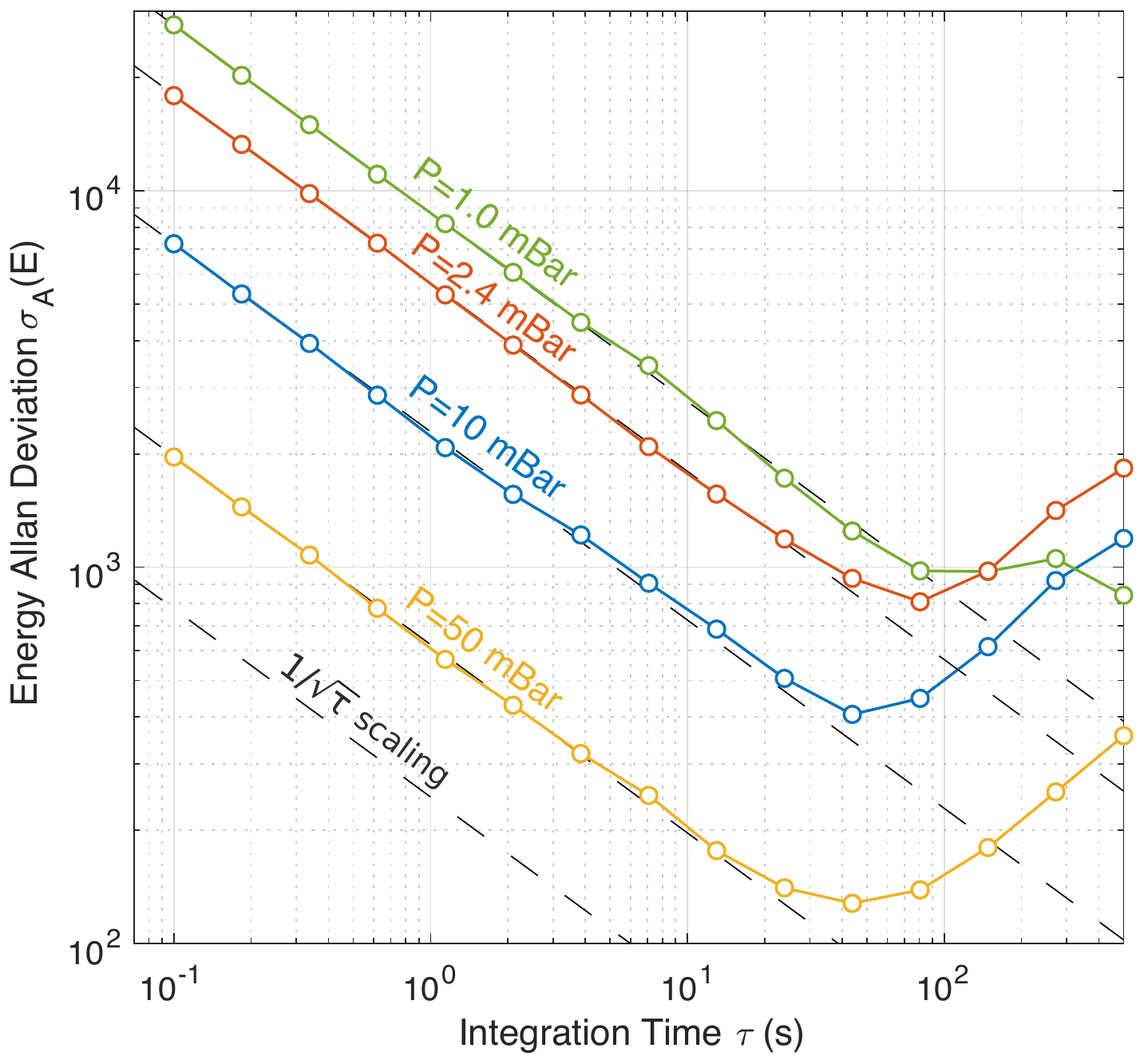}
\caption{\footnotesize \textbf{Energy Allan deviation measurement.} The Allan deviation of the energy $\hat{E} = \langle v^2 \rangle$ is shown as a function of the integration time $\mathcal{T}$, for different pressures.  At the operating pressure of $\sim 50\unit{mBar}$ we obtain maximum stability for $\mathcal{T} \simeq 40\unit{s}$. Lower pressures result in a higher Allan deviation and consequently a degraded stability, probably due to the onset of higher nonlinearities in the dynamics of the particle.}
\label{fig:S01_AllanDev}
\end{figure}

\vspace{0.3cm}
\noindent{\bf S2.\hspace{.2cm}Energy estimation} \mbox{}\\
The statistical nature of the errors affecting $S_v(\odr)$, $S_v^{\rm th}(\odr)$ and $\Gamma$ ideally allows one to increase the integration time for the measurement in order to arbitrarily reduce the associated errors $\sigma_{S_v}$, $\sigma_{S_v^{\rm th}}$, and $\sigma_{\Gamma}$. However, this approach is only valid as long as the system is not affected by slow drifts that affect the system over long timescales. As demonstrated by Hebestreit et al.~\citep{Hebestreit2017Calibration}, one  can define an optimal measurement time to extract the \emph{energy} $\hat{E}(\mathcal{T}) = \langle v^2 \rangle_{\mathcal{T}}$ of the particle. Here, $\mathcal{T}$ expresses the time span of the dataset from which the variance $\langle \cdot{} \rangle$ is calculated. Note that $\hat{E}$ is not the physical energy, but is proportional to it. More precisely we have $\hat{E} = \frac{2E_{\rm pot}}{m \Oo^2}$. The longest useful integration time is thus assessed through the \emph{Allan Deviation} of the variance $\hat{\sigma}_{E}$, calculated from a long position time trace $v(t)$ of $\sim 2\unit{hours}$. After chopping $v(t)$ into $N_{\mathcal{T}}$ shorter sections of variable length $\mathcal{T}$, we compute $\hat{E}_{j}(\mathcal{T})$ for each section $j=1,\hdots,N_{\mathcal{T}}$. The energy Allan deviation is then calculated for each value of $\mathcal{T}$ as:
\begin{equation*}
\hat{\sigma}_{E} (\mathcal{T}) = \sqrt{\frac{1}{2(N_{\mathcal{T}}-1)} \sum_{j=1}^{N_{\mathcal{T}}}\left(\hat{E}_{j+1}(\mathcal{T})-\hat{E}_j(\mathcal{T})\right)}
\end{equation*}
In figure~\ref{fig:S01_AllanDev} we show the \emph{Allan deviation} of the variance $\hat{\sigma}_{A}$, calculated for different pressures in the vacuum chamber. The experimental result demonstrate maximum stability for integration times of $\mathcal{T} \sim 40\unit{s}$. Moreover, for higher pressures we observe lower minima in the Allan deviation, which indicates a better stability of the system. We believe this is due to the influence of nonlinearities, that are minimized for pressures $P>10\unit{mBar}$.

\vspace{0.3cm}
\noindent{\bf S3.\hspace{.2cm}Net charge control} \mbox{}\\
The particles loaded in the optical trap usually show a non-null initial net charge $n_q^{\rm in}$ of the order of ten elementary charges. Statistics on the value of $n_q^{\rm in}$ shows a $70-30$ biased polarity distribution toward positive charges. However, higher flexibility on the particle's polarity and on  the number of charges $n_{q}$ to be used in the experiments is desirable. Moreover, this initial charge is not known a priori and an experimental method to measure the particle's charge is paramount for quantitative analysis of the electrically driven dynamics. To address these issues, the net charge of the levitated particles can be finely controlled in our system. Similarly to what has been shown in \citep{Frimmer2017Controlling}, we apply a high voltage to a bare electrode situated on a side of the vacuum chamber, few centimeters away from the optical trap. In a moderate vacuum of $P\sim 1\unit{mbar}$, the high voltage creates a plasma via the process of corona discharge~\citep{Fridman2004Plasma}. The polarity of the voltage sets the polarity of the corona, which ultimately allows one to bias the ratio of positive-to-negative ionized molecules that are accelerated toward the center of the chamber where the optical trap is situated. When adsorbed on the particle surface, these change its net charge almost monotonically (exception made for few random unfavorable events) from positive to negative and viceversa, depending on the polarity chosen. Disconnecting the high voltage stops the corona discharge and the number of charges on the particle is maintained stable indefinitely. 

\begin{figure}[h]
\includegraphics[width=\linewidth]{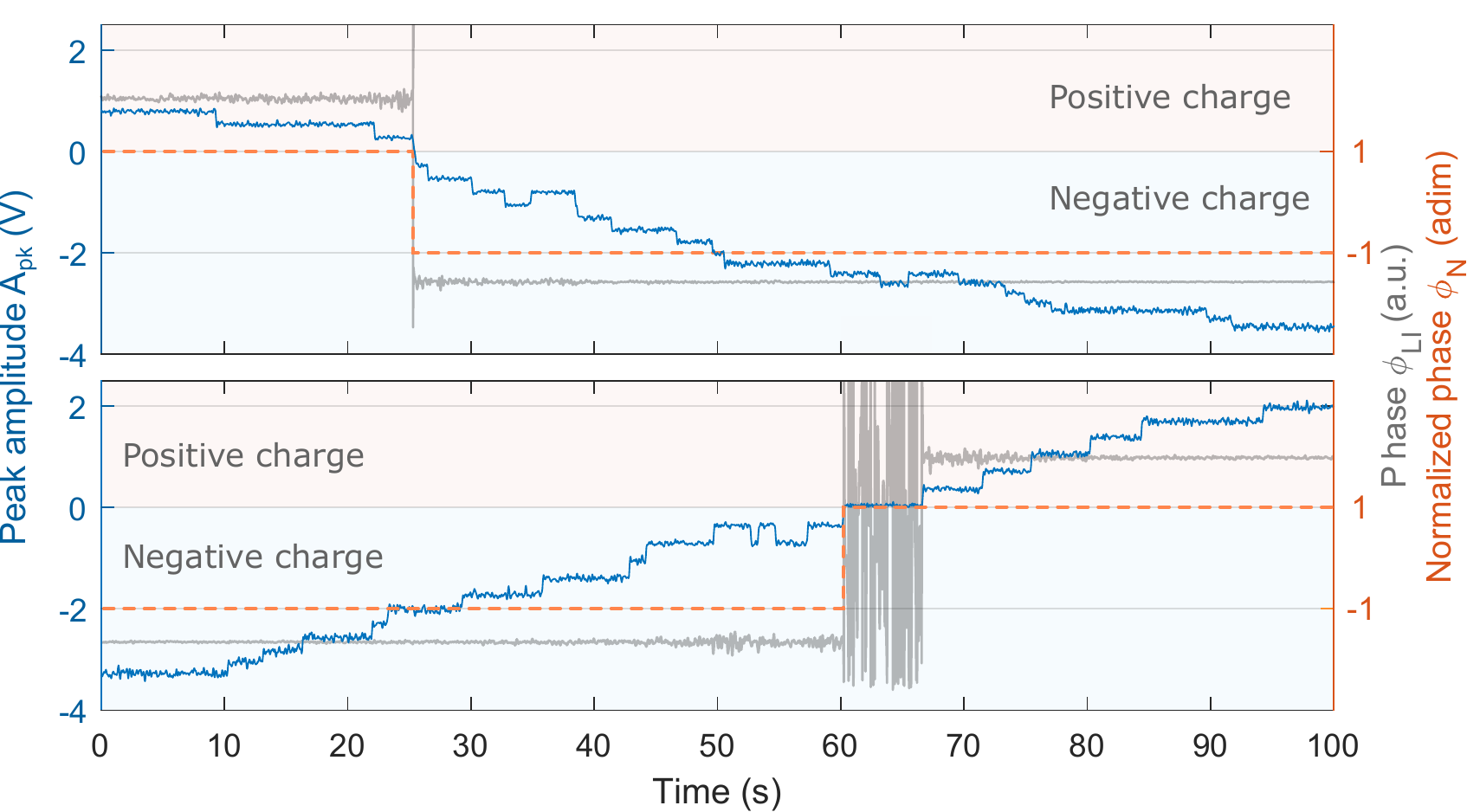}
\caption{\footnotesize \textbf{Net charge control. } In order to monitor the value of $n_q$ while the corona discharge is active, we use a lock-in amplifier to track in time the amplitude $A_{\rm LI}$ and phase $\phi_{\rm LI}$ of the driven oscillation. A normalized phase $\phi_{\rm N}$ is introduced to discern the polarity of the charged particle and the signed peak amplitude $A_{\rm pk}=A_{\rm LI} \cdot \phi_{\rm N}$ is reported in the two panels. Discrete steps are observed in $A_{\rm pk}$ identify single elementary charges being added  or removed from the particle.} 
\label{fig:S02_ChargeControl}
\end{figure}

To monitor changes in the particle's net charge while the corona is active we look for integer steps in the height of the peak $S_v(\odr)$ represented in Fig.~2 of main text. For this task we make use a lock-in amplifier which is synchronized with the function generator that drives the particle at frequency $\odr/2\pi$. As such the lock-in it provides the amplitude $A_{\rm LI}$ and phase $\phi_{\rm LI}$ of the driven oscillation, measured in a narrow band around $\odr$. It is important to stress that the lock-in only provides positive amplitudes, therefore making it impossible to discern the particle's polarity. In other words, the detected peak $S_v(\odr)$ takes the same value for both $\pm n_q$ and assumes the value of the thermal noise floor $S_v^{\rm th}(\odr)$ when $n_q = 0$. However, attraction/repulsion of Lorentz force is such that the particle's oscillation results in phase with the external driving if its polarity is positive, out of phase if this is negative and possesses an undefined phase if the particle is neutral. We can therefore introduce a normalized phase $\phi_{\rm N}$ of the form:
\begin{equation*} %\label{eq:03_NormPhase}
\left\{
\begin{alignedat}{3}
            &\phi_{\rm N} = 1,~&&\text{if}~~&&\phi_{\rm LI} > 0\\
         &\phi_{\rm N} = 1,~&&\text{if}~~&&\phi_{\rm LI}~~\text{is undefined} \\
     &\phi_{\rm N} = -1,~&&\text{if}~~&&\phi_{\rm LI} < 0~,
\end{alignedat}
\right.
\end{equation*}
such that the signed amplitude $A_{\rm pk} = A_{\rm LI} \cdot \phi_{\rm N}$ is now directly proportional to the charge (polarity included) of the particle. Figure \ref{fig:S02_ChargeControl} show $A_{\rm pk}$ (blue data), $\phi_{\rm LI}$ (gray data) and $\phi_{\rm N}$ (orange data) respectively for a discharging and charging process in the range $n_q \in [-16,6]$. Discrete steps in $A_{\rm pk}$ can indeed be observed, proving the reliability of our method of controlling the net charge down to the single elementary charge resolution.

\begin{figure}[h]
\includegraphics[width=\linewidth]{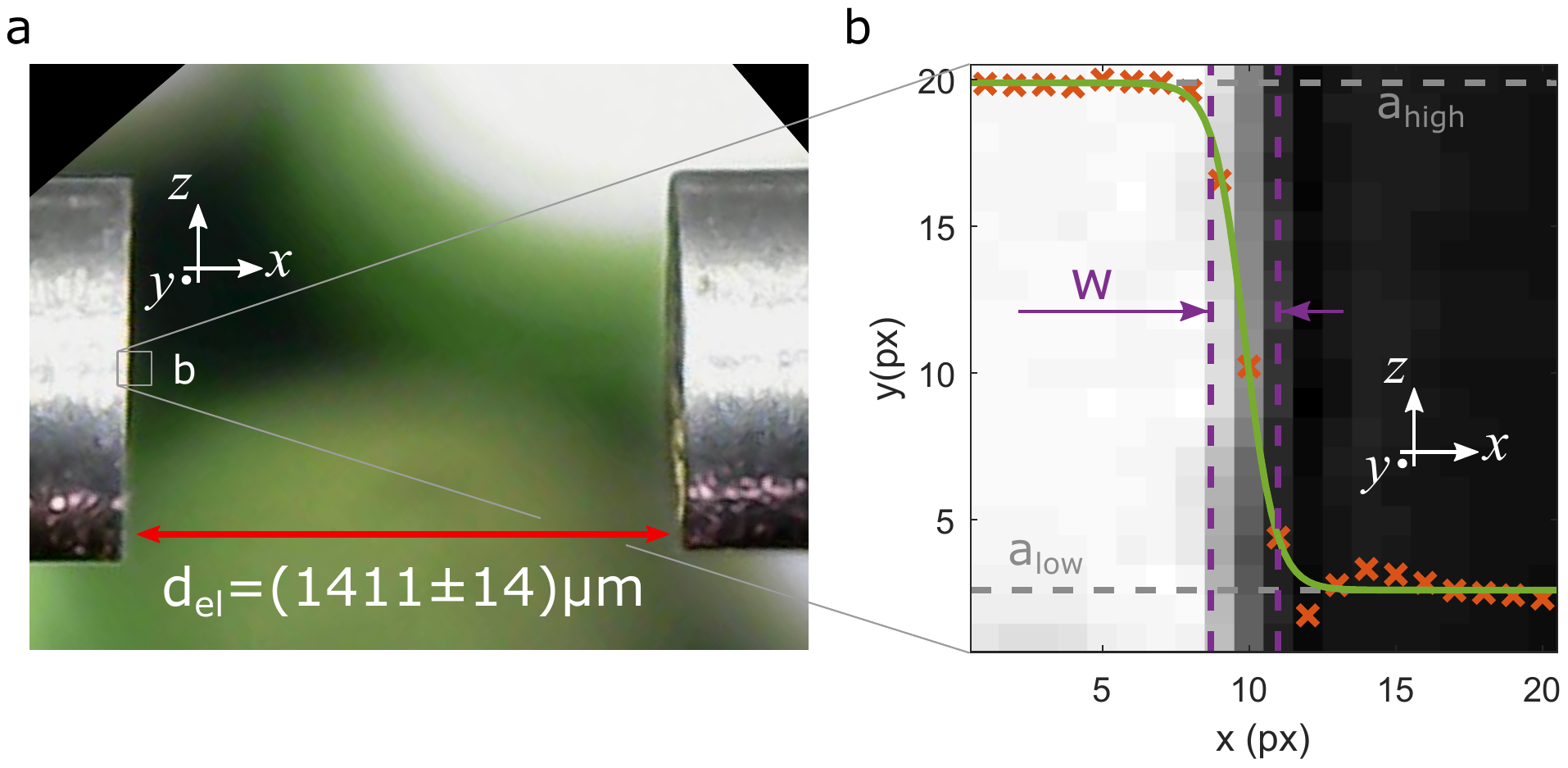}
\caption{\footnotesize \textbf{Electrodes geometry and position error estimation}. (a) A high resolution image of the electrodes geometry taken with a portable microscope that we fit directly into the vacuum chamber. Distances are calibrated via the electrodes diameter, which are known up to $1\unit{\mu m} $ tolerance. We measure a separation between the electrodes $d_{\rm el} = 1410\unit{\mu m}$ . (b) The uncertainty $\sigma_{d_{\rm el}}$ over the electrodes' gap is calculated from the edge blurring (b) of the image (see supplementary text for further details)}. 
\label{fig:S04_ElectrodeGeometry}
\end{figure}

\vspace{0.3cm}
\noindent{\bf S4.\hspace{.2cm} Electrodes geometry and electric field simulation}  \mbox{}\\
The electric force applied to the particle is of the form $F_0(t) = n_q \cdot q_e \cdot E_0 \cos(\odr t)$. In the present work $n_q$, $q_e$ and $\odr$ are assumed to be error free. On the other hand, the absolute value and the corresponding uncertainty on the electric field $E_0$ have a high impact on the measurement of the mass (we see indeed from Tab. 1 in the main text that the electric field provides the biggest relative error in our set of variables). In order to estimate the electric field we performed finite element calculations (COMSOL). The geometry of the system was inferred from  a high resolution image of the electrodes (see Fig.~\ref{fig:S04_ElectrodeGeometry}a), obtained \emph{in situ} with a portable microscope. The maximum achievable resolution was limited by the actual fitting of the microscope inside the vacuum chamber and by the field of view of the microscope that needed to include both electrodes in the same image. The factor $c_{\rm img} = 4.64 \pm 0.04 \unit{\mu m / px}$ used to calibrate pixels into physical distance units is calculated comparing the size of the electrodes with their nominal diameter: $\phi = 1\unit{mm}$, ISO h6 tolerance corresponding to $\sigma_\phi = 3\unit{\mu m}$. 

The uncertainty on the distance $d_{\rm el}$ separating the two electrodes is derived from the uncertainty on each electrode's edge position.We crop a $20 \times 20\unit{px^2}$ area at the edge of one of the electrodes, close to its center (see grayscale map in Fig.~\ref{fig:S04_ElectrodeGeometry}b). Averaging along the $z$ axis, provides the experimental profile $I(x)$ of the electrode (red crosses). A \emph{sigmoid} function of the form 
\begin{equation*}
\Sigma(x) =  a_{\rm high} - \frac{a_{\rm high}-a_{\rm low}}{1+\e{-(x-x_0)/\tau}}
\end{equation*}
is used to fit the profile, the result being displayed as a green solid line in Fig.~\ref{fig:S04_ElectrodeGeometry}b. The fitting parameters $a_{\rm high/low}$, $x_0$, $\tau$ representing respectively the high/low plateau level, the center of the edge, and its width are used to calculate the separation $w$ between $1/10$ and $9/10$ of the step height (dashed purple vertical lines). The error on the position of the single electrode edge is then $\sigma_p = w/2 = 1.14\unit{px}$, and the error on the electrode distance $\sigma_{\Delta p} = \sqrt{2} \sigma_p$. Using $c_{\rm img}$ to calibrate into physical units and to propagating with the corresponding error $\sigma_{c_{\rm img}}$, we finally obtain $d_{\rm el} = 1411 \pm 13 \unit{\mu m}$. The mapped geometry of the electrodes, together with their distance from the objective and from the collection lens are plugged into a COMSOL finite element calculation in order to estimate the electric field at the particle's position. The objective and the collection lens are both modeled as a metal holder with an inset dielectric. One of the two electrodes is grounded (i.e. $U=0\unit{V}$ potential), while on the other we apply a dc voltage ($U=1\unit{V}$).
In our simulations, we neglected possible charges at the objective lens surface. Although this possibility cannot be excluded a priori, we have estimated their effects and concluded that these fall well within the uncertainty of the electric field provided by the COMSOL simulations. In fact, the displacement of the particle due to these surface charges is bound by the Rayleigh range $z_0$ of the beam. Displacement higher than $z_0$ would indeed introduce systematic loss of the particle, which we do not observe. Therefore, being $z_0 \ll \sigma_{\rm d_{el}} = 14~\mu{\rm m}$, the effects of such charges introduce a disturbance on the electric field that is included already in the uncertainty $\sigma_{E_0}$ estimated.
Figure~\ref{fig:S04_ElectrodeGeometry} shows the outcome of the simulation. As expected, the effects of the collection lens are negligible, while we see that the field gets slightly affected by the objective and the dielectric lens. Nevertheless, in between the electrodes the field is homogeneous, and we find $E_0 = (577 \pm 6)\unit{V/m}$. The uncertainty on the value of $E_0$ is estimated performing several different simulations while varying the distance $d_{\rm el}$ and misplacing each of the electrodes by $\sigma_{d_{\rm el}} = 13\unit{\mu m}$. 
\begin{figure}[h]
\includegraphics[width=\linewidth]{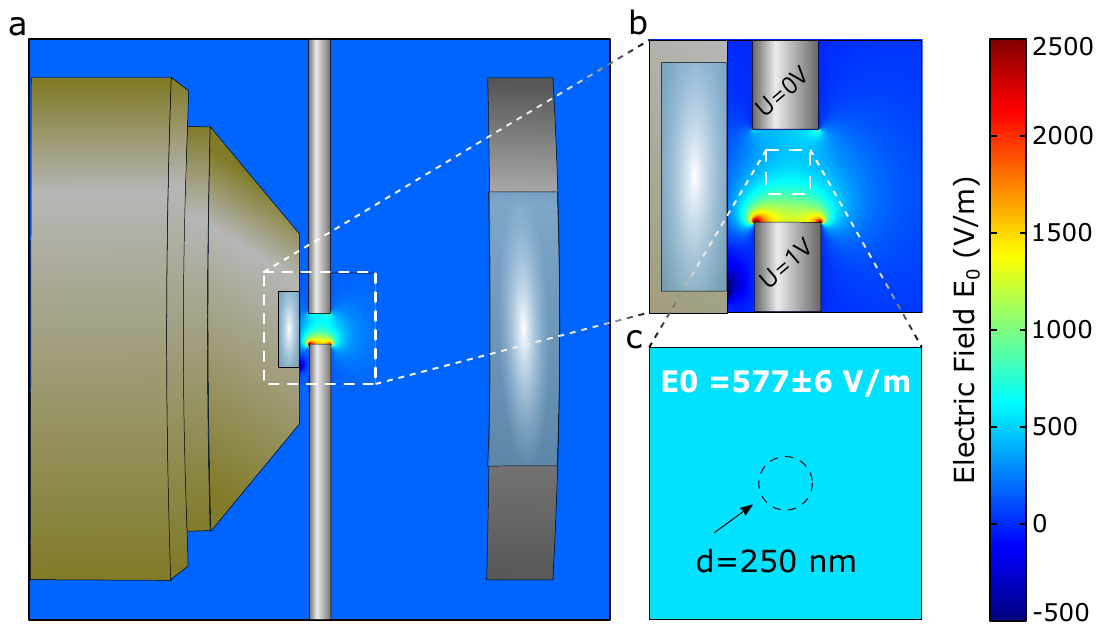}
\caption{\footnotesize \textbf{Electric field COMSOL simulation }. (a) Overall geometry of the set-up used to simulate the electric field at the trap position. (b) Magnified version of (a) shows that field between the electrodes is barely affected by the presence of the  dielectric lens inset in the objective. (c) The field $E_0$ can be considered constant and homogeneous in the volume explored by the particle. Uncertainty on the value of $E_0$ is calculated performing different simulations while varying the distance $d_{\rm el}$ and misplacing each of the electrodes by $\sigma_{d_{\rm el}} = 13\unit{\mu m}$}
\label{fig:S05_ElectricFieldSimulation}
\end{figure}	

\vspace{0.3cm}
\noindent{\bf S5.\hspace{.2cm}Simulation of a Duffing resonator} \mbox{}\\
With the estimated value of $\xi$, it was possible to simulate position time traces $x(t)$ in the presence and absence of non-linearities  and at different pressures. Comparing the results in the two cases allows us to obtain upper bounds on the errors of the fitted parameters: $\Omega_0$ and $\Gamma$. To this aim, we have performed Monte Carlo simulations with different values of $\xi \in [0, 12]\unit{\mu m^{-2}}$. Note that the upper bound is an overestimation of the actual Duffing value $\xi = (-9.03 \pm 0.44)\unit{\mu m ^{-2}}$ experimentally determined in Supplementary Section S1. This ensures that non-linearity based systematic errors on $\Gamma$ and $\Omega_0$ remain small enough, or even negligible.

The simulations are performed with a Runge-Kutta method of strong order 1\citep{Rossler2009Second}, which we detail in what follows: let $\mathbf{X}(t) \in \mathbb{R}^n$ be the stochastic process that we want to simulate, satisfying the general It\^o stochastic differential equation (SDE):
$$
\dif \textbf{X} = \textbf{a}(t, \textbf{X})\,\dif t+ \textbf{b}(t, \textbf{X})\,\dif W.
$$
Given a time step $\Delta t$ and the value $\textbf{X}(t_k)= \textbf{X}_k$, then $\textbf{X}(t_{k+1})$ is calculated recursively as
$$
\begin{array}{rl}
\textbf{K}_1 = & \textbf{a}(t_k, \textbf{X}_k) \Delta t  +(\Delta W_k-S_k\sqrt{\Delta t})\cdot \textbf{b}(t_k, \textbf{X}_k),
\\
\textbf{K}_2 = & \textbf{a}(t_{k+1}, \textbf{X}_k+ \textbf{K}_1) \Delta t \,+\\
& (\Delta W_k+S_k\sqrt{\Delta t}) \cdot \textbf{b}(t_{k+1}, \textbf{X}_k+\textbf{K}_1),\\
\textbf{X}_{k+1} = & \textbf{X}_k + \frac12(\textbf{K}_1 + \textbf{K}_2),
\end{array}
$$
where $\Delta W_k \sim \mathcal{N}(0, \Delta t)$, and $S_k = \pm 1$, having each probability 1/2.

As described in the main text, the equation of motion of the center of mass of the levitated nanoparticle is
\begin{align*}
\dif x_t & = v_t \dif t ~,\\
m\dif  v_t& = -\nabla \Psi(x_t) \dif t -m\Gamma v_t \dif t + \sigma \dif W_t + F_\text{el}(t)\dif t ~,
\end{align*}
where 
$$
\Psi(x) = m\Omega_0^2\left(\frac{x^2}{2} + \frac{\xi\cdot x^4}{4}\right)
$$
and higher terms of the series expansion of the optical potential have not been taken into consideration.

\begin{figure}[h]
\includegraphics[width=\linewidth]{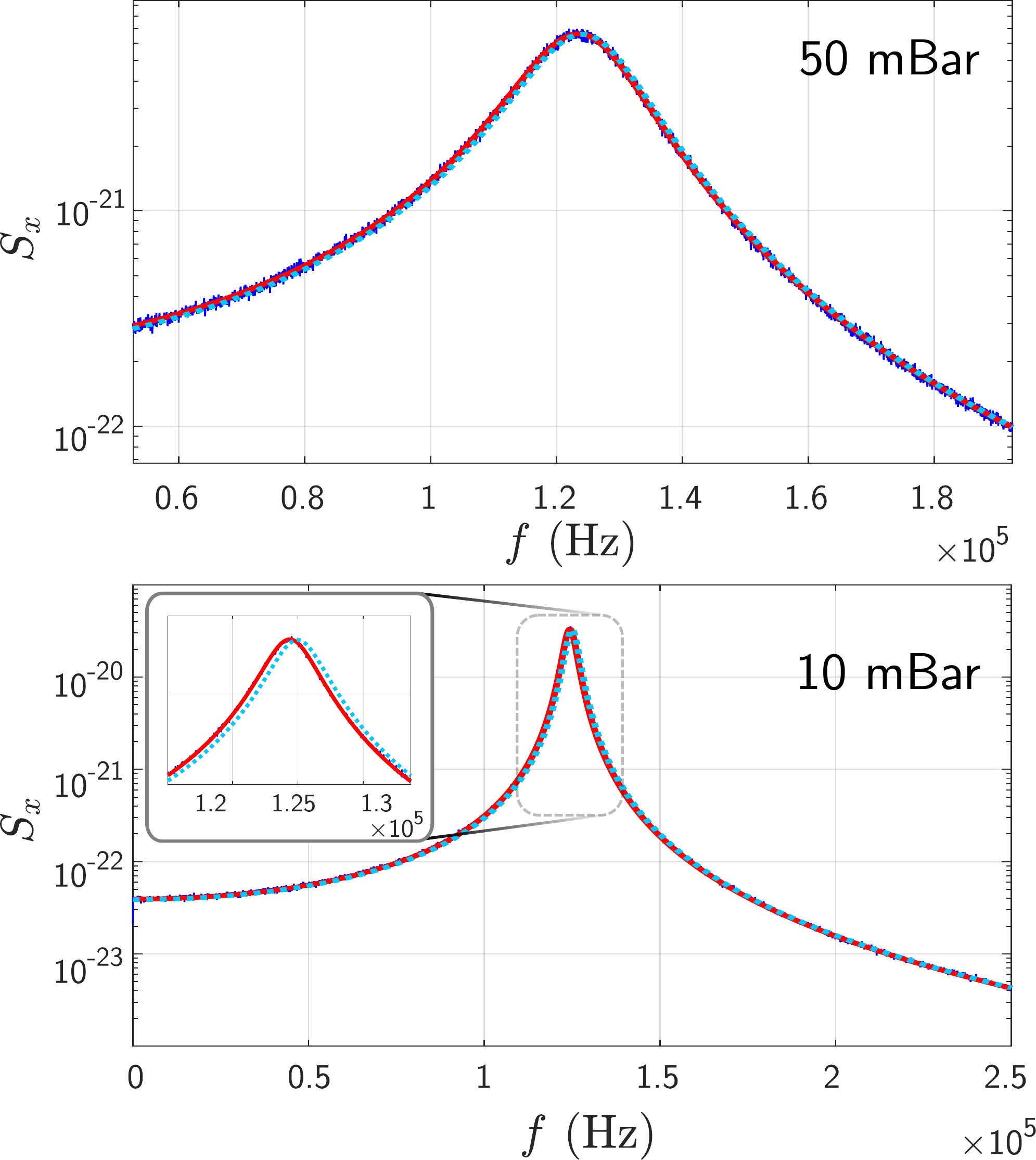}
\caption{\footnotesize \textbf{Estimated PSD from simulated traces}. Two PSDs, estimated from the simulation of 1000 traces of 40 ms each, are displayed, corresponding to pressures of $50\unit{mBar}$ and $10$10 mBar and 50 mBar. Dark blue corresponds to the results of the simulation, red corresponds to the fitted function (which assumes a linear model, even if the simulated potential has a Duffing term) and the discontinuous light blue corresponds to the exact linear response, calculated with the values of $\Gamma$ and $\Omega_0$ used in the simulation. As can be seen in the 10 mBar case, there is a shift in the resonance towards lower frequencies: since the resonance peak is higher the lower the value of $\Gamma$, this effect is more pronounced at lower pressures.}
\label{fig:S06_Simulation}
\end{figure}

The rest of the values needed to perform the simulations (i.e., $m$, $\Gamma$, $\sigma$ and $F_\text{el}$) have been calculated assuming:
\begin{itemize}
\item A temperature $T = 295$ K.
\item A spherical silica particle of radius $117.5$ nm and density 2200 kg/m$^3$.
\item $\gamma = m\Gamma$ follows Stoke's drag force, and is linear with the pressure for moderate levels of vacuum.
\item The noise intensity $\sigma$ satisfies the fluctuation-dissipation relation, i.e. $\sigma = \sqrt{2 k_B T m\Gamma}$.
\item The driving force $F_\text{el}$ obtained by the finite element method simulation of the electric field.
\end{itemize}

For a value of $\xi = 0$ and in the absence of electric driving, the equation of motion becomes a harmonic oscillator with additive stochastic driving, which is a particular case of the 2-dimensional OU process. As a last check to verify the results of the simulations, we have compared the estimated variance $\mathbb{E}[x^2(t)]$ with the analytical expression given by Wang et al.\citep{wang1945theory}. The results are in complete agreement.

To estimate the PSD of the center of mass motion $x(t)$ we used Bartlett's method with a rectangular window . $N_{\rm psd} = 1000$ PSDs, each one calculated as the periodogram of a $40\unit{ms}$ time trace, were averaged to obtain the final estimate of the spectral density $S_x(\omega)$. This PSD was later fitted with least squares to obtain the values of $\Omega_0$ and $\Gamma$. To estimate the errors and uncertainties in calculating $\Omega_0$ and $\Gamma$ with simulated data, we have used the exact same method, number of traces and time lengths.

The results of the simulations show that, at $P=50\unit{mBar}$, non-linearities result in relative errors on the estimated parameters of a few $0.1\%$. This is shown in table 1, main text, where an upper bound of the obtained values has been used.

For slightly lower pressures, however, the errors on the estimated parameters may become too large to provide accurate mass values. This is illustrated in figure \ref{fig:S06_Simulation}, where a comparison between the PSDs of traces at 10 mBar and 50 mBar is displayed. If the pressure at which the experiment is performed is even lower, the non-linearities can become comparable to the linear restoring force, and the proposed protocol for calculating the mass is not valid anymore.

\vspace{0.3cm}
\noindent{\bf S6.\hspace{.2cm}Spectral density derivation for the harmonic driving} \mbox{}\\
In the main text we use the non-unitary angular frequency Fourier transform, defined as
$$
G(\omega) = \mathcal{F}\left(g(t)\right) = \int_{-\infty}^\infty g(t)e^{-i\omega t}\mathrm{d}t.
$$
The power spectral density will then be 
\begin{align*} %\label{eq:psd}
\mathcal{S}_g(\omega) = \lim_{T \rightarrow \infty }\left| \frac{1}{\sqrt{T}}\int_{-T/2}^{T/2}g(t)e^{-i\omega t}\mathrm{d}t \right|^2,
\end{align*}
but since in any real experiment we only observe the dynamics of the system for a finite time $\mathcal{T}=2\tau$, we use the \emph{truncated} power spectral density, 
\begin{align}
S_g(\omega) &= \left| \frac{1}{\sqrt{\tau}}\int_{-\tau}^{\tau} g(t)e^{-i\omega t}\mathrm{d}t \right|^2 \nonumber  \\
&= \frac{1}{\tau}\left|\frac{1}{\pi}G(\omega) * \textrm{sinc}(\tau \omega)\right|^2. \label{eq:psd}
\end{align}
From linear time-invariant (LTI) theory we also know that if we apply a driving $f_\text{d}(t)$ (input) to a LTI system, the response (output) of the system will be 
$$
S_x(\omega) = |H(\omega)|^2|F_\text{d}(\omega)|^2
$$
where, in our case, $H(\omega)$ is the transfer function of the harmonic oscillator. Therefore, since $H(\omega)$ is known, to calculate $S_x(\omega)$ we only need to find an expression for $|F_\text{d}(\omega)|^2$. For a driving $f_\text{d}(t) = \cos \omega_\text{d} t$ and applying equation \eqref{eq:psd} we get
\begin{align*}
S_g(\omega) &= \frac{1}{\tau} \left|\frac{1}{\pi} \mathcal{F} (\cos \omega_\text{d} t) * \textrm{sinc}(\tau \omega)\right|^2 \\
& = \frac{1}{\tau} \left| \frac{1}{\pi}\left( \pi(\delta(\omega - \omega_\text{d}) + \delta(\omega + \omega_\text{d})\right) * \textrm{sinc}(\tau \omega)\cdot \tau\right|^2 \\
& = \frac{\tau}{2} \left| \textrm{sinc}[(\omega - \omega_\text{d})\tau] + \textrm{sinc}[(\omega + \omega_\text{d})\tau] \right|^2 \\
& \simeq \frac{\tau}{2} \textrm{sinc}^2[(\omega - \omega_\text{d})\tau].
\end{align*}
The error incurred in the last approximation is two orders of magnitude smaller than the uncertainties of the experiment, so we can safely assume that only the sinc function centered at $+\omega_\text{d}$ matters.

\begin{figure}
\includegraphics[width=\linewidth]{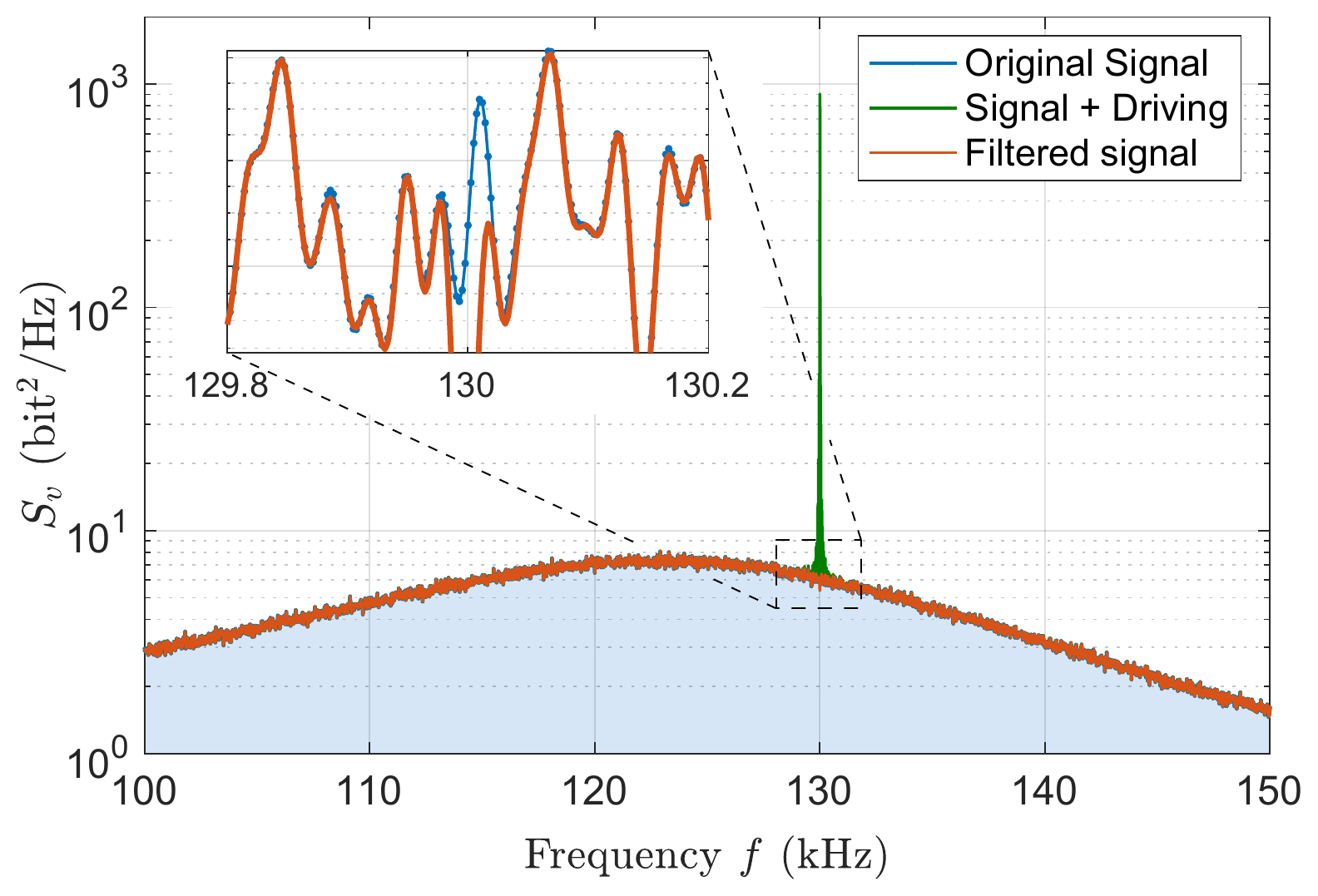}
\caption{\footnotesize \textbf{Peak filtering method}.}
\label{fig:S07_FilterPeak}
\end{figure}

\vspace{0.3cm}
\noindent{\bf S7.\hspace{.2cm}Peak filtering} \mbox{}\\
In the presence of electric driving $F_{\rm el}(t)$ the measured PSD $S_v(\omega)$ comprises a broad resonance $S_v^{\rm th}(\omega)$ (given by the thermal motion of the particle) and a narrow response peaking at the driving frequency $\odr$. In order to rigorously isolate the solely thermal contribution one should measure the particle's dynamics in the absence of the electric driving. However, to simplify the protocol it is possible to reconstruct and fit the thermal response by just filtering out the narrow peak at $\omega=\odr$. Here we demonstrate the validity of this procedure and estimate the corresponding error introduced in the estimation of the parameters $\omega_0$ and $\Gamma$. Figure~\ref{fig:S07_FilterPeak} exemplifies this method. A purely experimental thermal PSD (blue solid line) is numerically modified by adding in the time domain a coherent term of the form $F_0 \cos (\odr t)$ with a arbitrary amplitude $F_0$. Te corresponding PSD is shown as a solid green line. The inset of Fig.~\ref{fig:S07_FilterPeak} clearly shows that applying a notch filter of proper bandwidth ($2.6\unit{Hz}$ in the example shown) one can filter out the driving peak and obtain a filtered PSD that resembles the original data with a high degree of confidence. By independently fitting both curves, and comparing the results we obtain deviations on the order of $\sim 10^{-5}$. Likewise, provided the resonator is operated in the harmonic regime, the thermal energy $\langle v^2\rangle$ can be calculated from the filtered driven state with an associated error that remains below $\sim 0.02\%$.\\

\vspace{0.3cm}
\noindent{\bf S8.\hspace{.2cm}Temperature stability} \mbox{}\\

\begin{figure}[b]
\includegraphics[width=\linewidth]{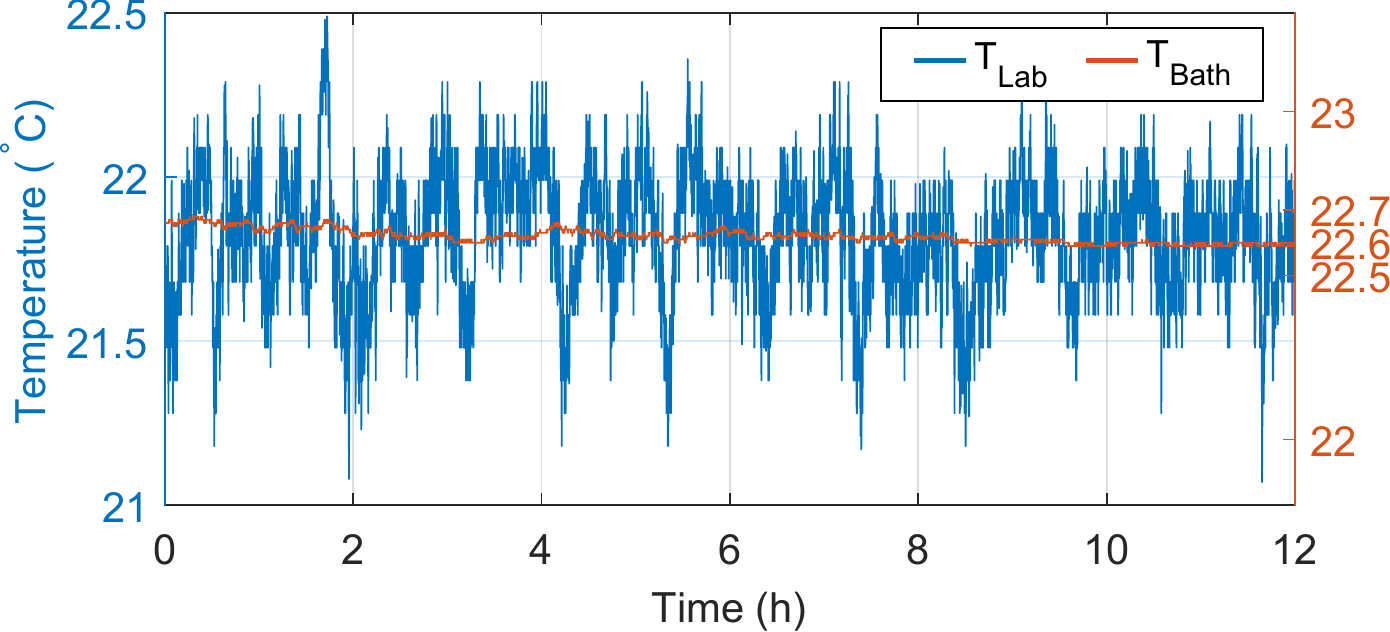}
\caption{\footnotesize \textbf{Temperature measurement in the system}. A first temperature probe (blue solid line) is placed in the middle of the lab, exposed to air turbulences and the constant variations due to the air conditioning. However, a second probe (red solid line) in contact with the vacuum chamber's walls measures the bath temperature surrounding the particle. A consistent reduction of the temperature fluctuations is achieved enclosing the set-up and carefully screening it from air turbulences.}
\label{fig:S8_temperature}
\end{figure}

The thermal bath surrounding the particle is assumed to be constantly thermalized with the set-up, and more precisely with the walls of the vacuum chamber, i.e. $T_{\rm bath} = T_{\rm chamber}$. Note that this assumption would not hold for low pressures ($P \lesssim 1\unit{mBar}$), where the rarified gas reduces the efficiency of convection and infrared laser absorption would rise the internal bulk temperature $T_{\rm bulk}$ of the particle~\cite{Hebestreit2018Measuring} and also of the trapping lens surface. In that case, one should include a two bath model~\cite{Millen2014Nanosacale} and consider a higher effective bath temperature and a corresponding higher uncertainty. However, the moderately high pressure $P= 50\unit{mBar}$ used in our measurements ensures the assumption $T_{\rm bulk} = T_{\rm bath} = T_{\rm chamber} = T$. To estimate this value, multiple temperature measurements on the surface of the vacuum chamber are carried out with a precision thermistor ($0.5^{\circ}{\rm C}$ accuracy) in order to exclude the presence of temperature gradients  and significant variations during the experimental times. Moreover COMSOL simulations of the objective lens absorption and gas convection at $P=50~{\rm mBar}$, estimate an increase of the dielectric material temperature of $\Delta T_{\rm lens} =3.6~{\rm mK}$, equivalent to $\Delta T_{\rm p} =3.1~{\rm mK}$ at the particle position. As a result, at these moderate pressures we can safely neglect temperature effects due to laser absorption, and the uncertainty on the bath temperature is therefore of the order of $\sigma_{T}/T \sim 0.2\%$. Note these effects would be much more critical at high vacuum ($P=10^{-6}~{\rm mBar}$)where a temperature increase of  $\Delta T \sim 15~{\rm K}$ was simulated under our typical experimental conditions. Finally, a stable and constant temperature $T$ is ensured only if the set-up is properly isolated from the lab environment. Isolation is achieved by enclosing the whole optical table in a unique box that maintains a more stable temperature and screens the setup from air turbulence. In fact, these would introduce pointing instabilities in the optical path, with the more drastic consequence being the unbalancing of the photoreceivers up to saturation. Figure~\ref{fig:S8_temperature} displays the measured temperature trend in the lab (probe placed in the middle of the room) and at the vacuum chamber position. We observe that while  the lab is exposed to temperature fluctuations of a few degrees, the temperature inside of the box screening the set-up is maintained stable, with variations that are well within the assumed accuracy.

\vspace{0.5cm}
\noindent{\bf S9.\hspace{.2cm} Error propagation} \mbox{}\\
All the quantities involved in the mass calculation are subject to both systematic and statistical errors. However, it is legitimate to assume that for $E_0$ and $T$, the systematic has a dominant contribution and the random error is negligible, while for $R_S$ and $\Gamma$ the opposite holds.

Moreover we can assume that the variables are uncorrelated, and we can therefore apply the well-known variance\citep{Ku1966Notes} formula to propagate the relative errors:
\begin{equation*} 
\sigma_{m} = \sqrt{\sum_i \left( \frac{\partial m}{\partial z_{i}}\right)^2 \sigma_{z_{i}}^2 },
\end{equation*}
where $z_i$ runs over the variables reported in table~1 of the main text, exceptions made for those in gray color for which $\sigma_{z_i} \simeq 0$.

As a result we obtain:
\begin{align}
\frac{\sigma_{m}^{\rm syst}}{m} &= \sqrt{\left( 2\frac{\sigma_{E_0}}{E_0}\right)^2 + \left( \frac{\sigma_{T}}{T}\right)^2 } \\
\frac{\sigma_{m}^{\rm stat}}{m} &= \sqrt{\left( \frac{\sigma_{S_v}}{S_v^{\rm el}}\right)^2 + \left( \frac{\sigma_{S_v^{\rm th}}}{S_v^{\rm el}}\right)^2 + \left( \frac{\sigma_{S_v^{\rm th}}}{S_v^{\rm th}}\right)^2 + \left( \frac{\sigma_{\Gamma}}{\Gamma}\right)^2}
\end{align}

Plugging in the errors reported in table~1, main text, we obtain $\sigma_{m}^{\rm syst}/m = 2.24\%$ and $\sigma_{m}^{\rm stat}/m = 0.91\%$.

\end{document}